\documentclass{ws-ijmpa}
\usepackage{xcolor}
\usepackage[sort,compress]{cite}
\usepackage[verbose]{hyperref}
\usepackage{dirtytalk}
\usepackage{soul}
\setstcolor{blue}
\hypersetup{colorlinks=false,allbordercolors=blue,pdfborderstyle={/S/U/W 1}}
\begin{document}

\markboth{Federico Scali}{An inductive approach to $f(R)$ gravity}

\catchline{}{}{}{}{}
%
%%%%%%%%%%%%%%%%%%%%%%%%%%%%%%%%%%%%%%%%%%%%%%%%%%%%%%%%%%%%%%%%%%%%

\title{An inductive approach to $f(R)$ gravity and its application to extended sources}

\author{Federico Scali\footnote{Via Valleggio 11, 22100, Como, Italy.}
}

\address{Department of Science and High Technology, University of Insubria, Via Valleggio 11\\
Como, 22100, Italy\\
INFN section of Milan, Via Celoria 16\\
Milan, 20133, Italy
\\
fscali@uninsubria.it}

\maketitle

\begin{history}
\received{Day Month Year}
\revised{Day Month Year}
\accepted{Day Month Year}
\published{Day Month Year}
\end{history}

\begin{abstract}
In a previous work (Phys. Rev. D 110, 064042), a new family of $f(R)$ gravitational Lagrangians was constructed inductively using only Solar System physics. Specifically, the modified Einstein equations were solved perturbatively on a Schwarzschild background, and the corresponding Lagrangian was reconstructed \textit{a posteriori} from the solution. Classical Solar System tests were then employed to constrain the fundamental length scale quantifying the deviation from general relativity.  In the first part of this work, the construction of the model is critically reviewed, emphasizing the generality of the approach and the applicability to different modified gravity frameworks. In the second part, the modified Newtonian potential produced by an extended source is derived within the new inductive $f(R)$ model, discussing the consistency of the Newtonian limit. In this context, the application to galactic dynamics emerges as an intriguing perspective.

\end{abstract}

\ccode{PACS numbers: 03.65.$-$w, 04.62.+v}
\clearpage
\section{Introduction}
Since its first formulation in 1915, Einstein's theory has received numerous experimental confirmations \cite{Will2014}. 
At the same time, in contexts different with respect to Solar System physics, the gravitational interaction lies at the centre of the greatest mysteries of contemporary physics. On the one hand, at the highest energies general relativity (GR) is difficult to reconcile with standard quantum field theory \cite{Giulini2003}; on the other, at length scales larger than Solar System's, Einstein's theory appears to require obscure forms of gravitating energies - namely dark matter and dark energy \cite{Gorini2024DM, Gorini2024DE} - to account for the dynamics of galaxies and the expansion of the universe \cite{Piattella2018}. 

All this suggests that the gravitational interaction is far from exhaustively understood, and opens the question of whether suitable effective modifications or extensions of Einstein's theory could explain these phenomena,\footnote{Namely, quantum effects and those associated to the the cosmological dark sector.} while preserving the well-tested Solar System and stellar-scale phenomenology of GR. In this view, Einstein's theory would be an effective description of the gravitational interaction, holding at the scales of the Solar System and of binary-stellar systems, but it could flow - with a suitable scale dependence - towards extended theories both at the microscopic and the cosmological scales.

Among the main avenues for modifying GR - see \cite{Saridakis2021ModifiedGravity, Nojiri2017, Clifton2012ModifiedGravity, Nojiri2011, Capozziello2010, DeFelice2010} for extensive reviews - a somewhat privileged role is played by the higher-derivative models, in which the free gravitational Lagrangian can be a general function of curvature invariants. On the one hand, this is because quantum effects are expected to produce higher order curvature corrections to the effective gravitational action \cite{Percacci2017}; on the other, it is because extensions of this type retain all the physical principles underlying GR, other than the strong equivalence principle \cite{Liberati2015, Will2014}.\footnote{Notice that this is non-trivially violated, since no additional gravitational field is introduced explicitly.} 

Since the vast majority of high-precision tests of gravity takes place in the Solar System and confirms GR with extreme accuracy, it is natural to expect that any extension eventually reduces to GR in the Solar System. This requirement is stronger than just compatibility with GR phenomenology, which does not imply it, and provides the perfect springboard to formulate the extension. In other words, while compatibility with observations constrains the parameters of a gravity theory, compatibility with GR local solutions can constrain the very structure of the theory. In the context of higher-derivative gravity models, the latter means constraining the dependence of the free gravitational Lagrangian from the curvature invariants.

In a recent work \cite{Scali2024}, this methodology was applied to $f(R)$ gravity, in which the gravitational Lagrangian is an arbitrary function of the scalar curvature only. Specifically, the modified Einstein equations for a spherically symmetric and static source were solved perturbatively, using scaling relations for the free functions that enforce compatibility with the usual Schwarzschild solution in the weak-field limit. The structure of the corresponding $f(R)$-gravitational Lagrangian was recovered \textit{a posteriori}, from the family of asymptotically-Schwarzschild solutions. The resulting theory depends on a fundamental length scale, parametrizing the deviation from GR, whose magnitude is constrained by the classical Solar System tests of gravity.

In the first part of this work, the construction of the model is reviewed, highlighting how the underlying procedure can be extended to different modified gravity frameworks. In the second part, the model is applied to extended sources in the Newtonian limit, from which the study of galactic dynamics emerges as an intriguing perspective. The latter is framed within an inductive route for modified gravity, progressing from the local Solar System scales to galaxies and ultimately to cosmology.
\section{An inductive approach to $f(R)$ gravity}
In \cite{Scali2024}, the simplest higher-derivative extension of Einstein's theory was considered, in which the free gravitational Lagrangian is generalized to an arbitrary and nonlinear function $f$ of the scalar curvature 
\begin{equation}
    S_f = \frac{1}{2k} \int_{\mathcal{M}} d^4 x \sqrt{-g} f(R).
    \label{eq: f(R) bulk action}
\end{equation}
Here $k = 8\pi G$ and $g$ is the determinant of the metric, whose signature is mostly plus. The integration is extended over the spacetime manifold $\mathcal{M}$. Neglecting boundary contributions,\footnote{If the domain of interest has a non-trivial boundary, $\mathcal{\partial M}$, then proper boundary conditions must be set on the dynamical fields to have a well-posed stationary action principle \cite{Dyer2008}. In GR, fixing the metric on the boundary is a necessary and sufficient condition, $\delta g_{\mu\nu}|_\mathcal{\partial M} = 0$, provided that the so-called Gibbons-Hawking-York boundary term is added to the free action \cite{Poisson2009}. In $f(R)$ gravity things are complicated by the non-linearity of the $f$, and specifying only the metric on the boundary is no more sufficient to make the gravitational action stationary on the solutions of (\ref{eq: Einstein equations for f(R)}). In the present context, however, the local solutions of Eqs.\ (\ref{eq: Einstein equations for f(R)}) for the gravitational field in the Solar System are of interest. The boundary can thus be pushed sufficiently far from the source, where no generality is lost in setting the scalar curvature to vanish, as an additional boundary condition ($\delta R|_\mathcal{\partial M} = 0$). In this case, the action principle is made well-posed by a simple extension of the GHY term of GR \cite{Dyer2008}.} 
variation of $S_f$ plus a matter action $S_m$ provides a modified set of Einstein equations
\begin{equation}
    \phi R_{\mu\nu} - \frac{1}{2}g_{\mu\nu}f(R) + g_{\mu\nu}\nabla^2 \phi - \nabla_\mu\nabla_\nu \phi = k T_{\mu\nu},
    \label{eq: Einstein equations for f(R)}
\end{equation}
where $\phi \equiv \frac{\partial f(R)}{\partial R}$ and $T^{\mu\nu} \equiv \frac{2}{\sqrt{-g}}\frac{\delta S_m}{\delta g_{\mu\nu}}$ is the energy-momentum tensor of matter. If $\phi$ is not constant, the equations are fourth order in the derivatives of the metric. 

In the spirit of the introduction, the aim is to constrain the form of the $f(R)$ by using Solar System physics, particularly by requiring compatibility with the GR solution in this setting. With good approximation, the Solar System is made of test bodies orbiting around a spherically symmetric and static source, whose gravitational field is 
\begin{equation}
 ds^2 = -e^{\nu(r)}dt^2 + e^{\mu(r)}dr^2 + r^2 d\Omega^2,
 \label{eq: line element of a spherical system in proper coordinates}
\end{equation}
in Schwarzschild coordinates. The symmetries of the configuration reduce Eqs.\ (\ref{eq: Einstein equations for f(R)}) to a system of three equations of second order in $\mu,\nu$ and $\phi$ \cite{Scali2024}. 
Compatibility with GR requires that the solution of (\ref{eq: Einstein equations for f(R)}) does not differ significantly from the usual Schwarzschild line element at the location of the high-precision experiments, that is, many Schwarzschild radii away from the source, where the gravitational field is weak. In formulae 
\begin{equation}
    \begin{aligned}
        &\phi(r) = 1+\sigma(r),\\
        &\nu(r) = \ln\left(1-\frac{2M_\odot G}{r}\right) + g(r),\\
        &\mu(r) = - \ln\left(1-\frac{2M_\odot G}{r}\right) + m(r),
    \end{aligned}
    \label{eq: Schwarzschild potentials with arbitrary corrections}
\end{equation}
where $\sigma,g$ and $m$ are subleading in the limit $r \gg 2M_\odot G$. On this note, it is natural to assume for $\sigma,g$ and $m$ the same expansions holding for the Schwarzschild counterparts in the weak-field limit 
\begin{equation}
    \sigma(r) = \sum_{i=1}^{+\infty} \frac{\alpha^{\sigma}_i}{r^i},\,\,\,
    g(r) = \sum_{i=2}^{+\infty} \frac{\alpha^{g}_i}{r^i},\,\,\,
    m(r) = \sum_{i=2}^{+\infty} \frac{\alpha^{m}_i}{r^i}.
    \label{eq: Taylor-Laurent expansion of the corrections to the Schwarzschild potentials}
\end{equation}
This choice restricts the space of possible solutions,\footnote{For example, the expansion (\ref{eq: Taylor-Laurent expansion of the corrections to the Schwarzschild potentials}) does not account for Yukawa-like corrections \cite{Capozziello2010}, which are usually found under the assumption of analytical $f(R)$ \cite{AparicioResco2026}. Nonetheless, notice how Yukawa-like corrections are structurally subleading in the asymptotic region with respect to the class of corrections in Eqs.\ (\ref{eq: Taylor-Laurent expansion of the corrections to the Schwarzschild potentials}).} but it allows a direct comparison with the Schwarzschild potentials in the same limit and, importantly, it allows to approach perturbatively the solution of (\ref{eq: Einstein equations for f(R)}). In particular, given any order $n$ for the leading term in the expansion for $\sigma$, the Eqs.\ (\ref{eq: Einstein equations for f(R)}) imply \cite{Scali2024}
\begin{equation}
    \begin{aligned}
        &\phi(r) = 1 + \frac{c_1}{r^n} +\mathcal{O}\left(\frac{1}{r^{n+1}}\right) ,\\
        &\nu(r) =  \ln\left(1-\frac{2M_\odot G}{r}\right) - \frac{c_1}{r^n} +\mathcal{O}\left(\frac{1}{r^{n+1}}\right),\\
        &\mu(r) = -  \ln\left(1-\frac{2M_\odot G}{r}\right) - n\frac{c_1}{r^n} +\mathcal{O}\left(\frac{1}{r^{n+1}}\right),\\
        &n\geq 2, 
        \label{eq:Leading order corrections to the scalar and the Schwarzschild potentials}
    \end{aligned}
\end{equation}
where the first non-zero constant in the expansion for $\sigma$ has been renamed $c_1$. The resulting modified Schwarzschild line element is 
\begin{multline}
    ds^2 = 
    - \left(1- \frac{2M_\odot G}{r}\right) dt^2  
    +\frac{1}{1-\frac{2M_\odot G}{r}} dr^2 + r^2d\Omega^2 \\
    - \frac{c_1}{r^n}(-dt^2 + ndr^2) + \mathcal{O}\left(\frac{1}{r^{n+1}}\right).
    \label{eq: asymptotically Schwrzschild solutions}
\end{multline}
The $f(R)$ gravitational Lagrangian can be reconstructed \textit{ex-post} from the solution (\ref{eq: asymptotically Schwrzschild solutions}). Plainly, by computing the scalar curvature of (\ref{eq: asymptotically Schwrzschild solutions}) at leading order 
\begin{equation}
    R = 3n (n-1)\frac{c_1}{r^{n+2}} + \mathcal{O}\left(\frac{1
    }{r^{n+3}}\right),
    \label{eq: leading correction to the scalar curvature computed from the metric}
\end{equation}
and by recalling that $\phi \equiv \frac{\partial f(R)}{\partial R}$, integration of $\phi$ gives
\begin{equation}
    f(R) = R + \frac{1}{2}|c_1|^{\frac{2}{n+2}}\frac{n+2}{(n+1)(3n^2-3n)^{\frac{n}{n+2}}}|R|^{2\frac{n+1}{n+2}},
    \label{eq: a posteriori recovered form of the leading correction to Einstein-Hilbert action}
\end{equation}
where the constant of integration is set to vanish, since GR must be recovered when $c_1 = 0$, and higher orders in the curvature are intended. 

To sum up, compatibility with GR at the local scales - expressed through Eqs.\ (\ref{eq: Schwarzschild potentials with arbitrary corrections}, \ref{eq: Taylor-Laurent expansion of the corrections to the Schwarzschild potentials}) - made it possible to recover perturbatively the one-parameter family of asymptotically Schwarzschild solutions (\ref{eq: asymptotically Schwrzschild solutions}) and, in turn, the corresponding family of gravitational Lagrangians (\ref{eq: a posteriori recovered form of the leading correction to Einstein-Hilbert action}). The dependence from the order parameter $n$ tells that the sole requirement of compatibility with GR in the weak-field limit leaves some freedom in the structure of the Lagrangian. The latter turns out to be non-analytic at the background value $R = 0$, since the exponent in (\ref{eq: a posteriori recovered form of the leading correction to Einstein-Hilbert action}) is always rational and less than quadratic. This case is somewhat complementary with respect to the usual scenario in which the $f(R)$ can be Taylor expanded \cite{AparicioResco2026}.

Non-analytic $f(R)$ models are not new in the literature: examples trace back to the first inverse curvature models \cite{Carroll2004CosmicSpeedUp}, to inverse power law models \cite{Nojiri2003}, to fractional power laws \cite{Lecian2009} and logaritmic corrections \cite{Capozziello2016Mass-radius} (see \cite{Nojiri2017,Nojiri2011,DeFelice2010} for extensive reviews). Therefore, although non-analyticity may introduce pathologies in the theory (e.g.\ instabilities), it is generally viewed as a possible feature of an $f(R)$ model \cite{DeFelice2010}. A relevant comparison for the present case is the non-analytic power-law model, $f(R) = R + \gamma R^\beta,\, 2<\beta<3$, proposed by Lecian and Montani \cite{Lecian2009} as a viable theory on astrophysical scales. Besides the parameters space,\footnote{The exponent in (\ref{eq: a posteriori recovered form of the leading correction to Einstein-Hilbert action}) takes values in the semi-open set $\left[\frac{3}{2},2\right)$ and it is always rational.} the main difference is that the model (\ref{eq: a posteriori recovered form of the leading correction to Einstein-Hilbert action}) is recovered \textit{a posteriori} from a local solution rather than postulated, which is closer in spirit to some reconstructive methodologies used in a cosmological context \cite{Nojiri2007ModifiedGravityAndItsReconstruction,Capozziello2014ConnectingEarlyAndLateUniverseWithfR}.

As a result of such inductive procedure, the Lagrangian (\ref{eq: a posteriori recovered form of the leading correction to Einstein-Hilbert action}) is on-shell with respect to the solutions (\ref{eq: asymptotically Schwrzschild solutions}); hence, it is strictly valid on the local scales of the Solar System\footnote{A deeper discussion on this point can be found in \cite{Scali2024}.} and depends on the ansatz (\ref{eq: Taylor-Laurent expansion of the corrections to the Schwarzschild potentials}). It is important to remark that Eq.\ (\ref{eq: a posteriori recovered form of the leading correction to Einstein-Hilbert action}) should be interpreted as the local limit of an exact $f(R)$ model, since it is reconstructed from a perturbative solution of the $f(R)$ equations. Also, notice that $f'(R)$ is positive and finite in the low curvature limit, hence the theory is ghost-free in its scope of validity \cite{Saridakis2021ModifiedGravity}.\footnote{If the Lagrangian (\ref{eq: a posteriori recovered form of the leading correction to Einstein-Hilbert action}) were promoted to an exact model, the sign of the first derivative would depend on the sign and magnitude of the scalar curvature.} The second derivative of (\ref{eq: a posteriori recovered form of the leading correction to Einstein-Hilbert action}) is instead strictly positive for any value of $R$, hence the theory does not present Dolgov-Kawasaki instability either \cite{DolgovKawasaki2003,Faraoni2006MatterInstability}. Inevitably, both these considerations hold at local scales, since the exact Lagrangian is not known a priori.\footnote{In this view, the stability requirements can be used to constrain the exact $f(R)$ Lagrangian, together with the consistency condition (\ref{eq: a posteriori recovered form of the leading correction to Einstein-Hilbert action}) on the local scales.}

The real constant $c_1$ in Eq.\ (\ref{eq: asymptotically Schwrzschild solutions}) quantifies the deviation from the Schwarzschild solution and appears in the gravitational Lagrangian (\ref{eq: a posteriori recovered form of the leading correction to Einstein-Hilbert action}); therefore, such constant is linked to the fundamental length scale of deviation from GR. The latter must be constrained using Solar System observations, as shown in \cite{Scali2024}. Notice that, at this level, $c_1$ is somewhat phenomenological, since it is not motivated by fundamental arguments, and it is strictly tied to Solar System observations. Consequently, even if the central mass is removed ($M_\odot \to 0 $) the constant $c_1$ still characterizes the solution; however, observational tests such as the gravitational lensing would constrain its magnitude to vanish.

The correspondence of the $f(R)$ model (\ref{eq: a posteriori recovered form of the leading correction to Einstein-Hilbert action}) with scalar-tensor theory - see for example \cite{Capozziello2009} - was also studied in \cite{Scali2024}, with the aim to provide a more rigorous definition of the length scale of deviation from GR. It was argued how the equivalence in general depends on the solution of $F(R,\chi) = f''(\chi)(R-\chi) = 0$, the field $\chi$ being the additional scalar. Specifically, if $f''$ is non-zero and sufficiently regular, then the equation is solved by setting $\chi(R) = R$ and the $f(R)$ equations are recovered. In the case of a non-analytic $f(R)$ this is a sensible point, because $f''$ may be non-regular. For the model (\ref{eq: a posteriori recovered form of the leading correction to Einstein-Hilbert action}) and the critical value $n = 2$ one finds $F \propto \frac{R-\chi}{\sqrt{|\chi|}}$; therefore, as long as $\chi \neq 0$, the zeroes of $F$ are defined by $\chi(R) = R$. One can realize, however, that $F(R,\chi)$ admits no limit for $(\chi, R)\to(0,0)$,\footnote{For example, by taking the limit along $R = \chi$ and $R = \chi + a\sqrt{|\chi|}$.} so that the equivalence would hold at most for $R$ in the open set $\mathbb{R}\setminus\{0\}$. One could \say{remove the singularity} induced by the defining equation and naturally set $\chi(0) \equiv 0$, thus enabling the usual scalar-tensor equivalence globally; however, in this case the equivalence would not be simply implied by the field equations.\footnote{This is a sensible mathematical point, whose implications would deserve a more thorough investigation.}

Letting this point aside, in \cite{Scali2024} is shown that for an $f(R)$ model of the type $f(R) = R + \alpha R^k$,\footnote{The model in (\ref{eq: a posteriori recovered form of the leading correction to Einstein-Hilbert action}) belongs to this family for $\frac{3}{2} \leq k < 2$.} with $\alpha > 0$ and $k >1$, the effective potential regulating the dynamics of the (redefined) scalar in the corresponding scalar-tensor theory reads $\left(\varphi \equiv \frac{df(\chi)}{d\chi}\right)$
\begin{equation*}
    V_{eff}(\varphi) = \frac{\alpha}{3}\frac{k-1}{2k-1}\left(\frac{|\varphi-1|}{\alpha k}\right)^{\frac{k}{k-1}}[(2-k)\varphi +3(k-1)].
\end{equation*}
Unless $\frac{k}{k-1}$ is an even integer - which never happens for the model (\ref{eq: a posteriori recovered form of the leading correction to Einstein-Hilbert action}) - the potential is not analytical at its (relative) minimum $\phi = 1$ and an effective mass cannot be defined in the usual way.\footnote{The square effective mass is provided by the coefficient of the quadratic term in the Taylor expansion of the effective potential around its minimum \cite{Faraoni2006MatterInstability}.} Also, notice that $V_{eff}$ is unbounded from below for every $k \neq 2$, leading to the so-called Ostrogradsky instability \cite{Woodard2007AvoidingDarkEnergy}. However, given the local and asymptotic nature of the model (\ref{eq: a posteriori recovered form of the leading correction to Einstein-Hilbert action}), in the present case the shape of the potential cannot be trusted far from the minimum. A possible Ostrogradsky instability should then be checked in the exact theory, reducing to (\ref{eq: a posteriori recovered form of the leading correction to Einstein-Hilbert action}) in the local limit.

Finally, the procedure presented here can be generalized to a five-steps constructive recipe for (Lagrangian) modified gravity models from Solar System physics. Namely 
\begin{itemize}
    \item consider an extension of the GR Lagrangian for matter and gravity, involving a number of unspecified functions - e.g.\ a function of more general curvature invariants - which can also be tensor valued, and derive the corresponding modified Einstein equations. 
    \item Impose compatibility with GR in the Solar System, by restricting the space of solutions to slight deviations from Schwarzschild.
    \item With the help of proper scaling relations for the free functions, solve perturbatively the Einstein equations. 
    \item Reconstruct \textit{a posteriori} the free functions in the Lagrangian for matter and gravity. 
    \item Constrain the free parameters of the perturbative solution with the high-precision Solar System observations. 
\end{itemize}
In principle, this procedure can be applied to different modified gravity frameworks - other than $f(R)$ gravity - and even to different physical contexts in which GR is well tested - other than the Solar System.

\section{Newtonian limit}
The interaction of a non-relativistic test particle moving in a weak gravitational field is well described by the coupling with a classical Newtonian potential $\Phi$. This is linked to the time-time component of the metric field by the relation\footnote{The correspondence (\ref{eq: correspondence between the time-time component of the metric and the classical gravitational potential}) results from requiring that the geodesic equation of a test body reduces to the classical equation of motion $ \frac{d^2\vec x}{dt^2} = - \vec \nabla \phi$, see \cite{Misner2017}.} \cite{Misner2017}
\begin{equation}
    g_{00} = -1 - 2\Phi. 
    \label{eq: correspondence between the time-time component of the metric and the classical gravitational potential}
\end{equation}
Therefore, within the theory developed so far - Eq.\ (\ref{eq: asymptotically Schwrzschild solutions}) - the modified Newtonian potential sufficiently far from a spherically symmetric and static source like the Sun reads 
\begin{equation}
    \Phi_{S^2}(r) = - \frac{MG}{r} - \frac{1}{2}\frac{c_1}{r^n} + \mathcal{O}\left(\frac{1}{r^{n+1}}\right), 
    \label{eq: classical potential for spherical body derived from the modified S solution}
\end{equation}
It is instructive to recover the same result from direct inspection of timelike geodesics in the metric (\ref{eq: asymptotically Schwrzschild solutions}). In particular, the existence of two (commuting) Killing vector fields, $\xi_t = \frac{\partial}{\partial t}$ and $\xi_\varphi = \frac{\partial}{\partial \varphi}$, implies that the quantities
\begin{equation}
\begin{aligned}
    \frac{E}{m} &= -g_{\mu\nu}u^\mu\xi_t^\nu = \left[1- \frac{2MG}{r} - \frac{c_1}{r^n} + \mathcal{O}\left(\frac{1}{r^{n+1}}\right)\right] \frac{d t}{d\tau},\\
        \frac{L}{m} &= g_{\mu\nu}u^\mu\xi^\nu_{\varphi} = r^2 \frac{d \varphi}{d\tau},
\end{aligned}
\end{equation}
are conserved along a geodesic. These are associated to the relativistic energy and angular momentum of the particle, whose four-velocity is $u^\mu = \frac{d x^\mu(\tau)}{d\tau}$. The normalization of $u^\mu$ then gives 
\begin{equation}
        \frac{E^2-m^2}{2m} = \frac{1}{2}m\dot r^2 + \frac{L^2}{2mr^2}  - \frac{mMG}{r} - \frac{MG L^2}{m r^3} - \frac{mc_1}{2r^n},
\end{equation}
where orders higher than $\mathcal{O}\left(\frac{1}{r^n}\right)$ have been omitted.\footnote{The expression would also involve a term $\propto \dot r^2 \frac{c_1}{r^n}$, which is subleading for slowly moving particles or quasi-circular orbits and can be neglected.} Hence, the system is equivalent to a test particle moving in one dimension under the influence of the effective potential 
\begin{equation}
    V_{eff}(r) =  \frac{L^2}{2mr^2} - \frac{mMG}{r} - \frac{MG L^2}{mr^3} - \frac{mc_1}{2r^n} + \mathcal{O}\left(\frac{1}{r^{n+1}}\right).
\end{equation}
Letting aside the centrifugal barrier, with respect to Eq.\ (\ref{eq: classical potential for spherical body derived from the modified S solution}) the additional term $\propto \frac{L^2}{r^3}$ is present, which in the usual Schwarzschild solution is the relativistic correction to the Newtonian potential \cite{Wald1984}. This term can compete with the modification $\sim \frac{c_1}{r^n}$ and would spoil the correspondence in Eq.\ (\ref{eq: correspondence between the time-time component of the metric and the classical gravitational potential}); therefore, to interpret the $\sim\frac{c_1}{r^n}$ term as the leading correction to the Newtonian potential, the analysis must be restricted to test particles with angular momentum $L^2 \ll \frac{m^2c_1}{MGr^{n-3}}$, with respect to the local source, for every $r$ along their trajectories. 

This requirement sets a consistency upper bound on the relative velocity of particles with respect to the central star. By considering circular orbits and by making use of the upper bounds on $c_1$ from the red-shift measurements \cite{Scali2024}, one finds $v^2 \ll 2 \left(\frac{2M_\odot G}{R}\right)^{n-1} (\alpha_n)^n$, where $v$ is the velocity, $R$ is the radius of the orbit and $\alpha_n$ stands for the upper bound on $\sqrt[n]{\frac{c_1}{(2M_\odot G)^n}}$.\footnote{See Table 1 in \cite{Scali2024}.} By inserting the $\alpha_n$'s and using the Sun-Earth distance as a proxy, it is found $v^{n = 2}_{max} \sim 3.8\times10^{-3}c$; $v^{n = 3}_{max} \sim 2.6\times10^{-4}c$, $v^{n = 4}_{max} \sim 1.8\times10^{-5}c$. Already for $n = 4$, the maximum velocity is smaller than the orbital velocity of the Earth around the Sun $\sim 30 \text{km/s}$. Therefore, the bound effectively restricts the applicability of the Newtonian limit (\ref{eq: classical potential for spherical body derived from the modified S solution}) - before the relativistic correction becomes comparable with the $f(R)$-induced correction - and may select only the first values of the parameter $n$.

\section{Extended sources}
Since Eq.\ (\ref{eq: classical potential for spherical body derived from the modified S solution}) holds for a spherically symmetric source like the Sun, it is natural to interpret it as the modified Newtonian potential produced by a point-like particle \cite{Capozziello2007LowSurfaceBrightness}, from the point of view of an extended source like a galaxy. Therefore, the (weak) potential of an extended source with volume density $\rho_V(\vec x)$ just results from the sum of all the individual point-particle contributions
\begin{equation}
    \Phi(\vec x) = -MG \int d^3\vec x' \frac{\rho_V(\vec x')}{|\vec x- \vec x'|} - \frac{c_1}{2} \int d^3\vec x' \frac{\rho_V(\vec x')}{|\vec x- \vec x'|^n},
    \label{eq: modified Newtonian potential generated by an extended source}
\end{equation}
since the superposition principle holds in the Newtonian limit. Here $M$ is the total mass of the extended source and the density $\rho_V$ is normalized to unity. 
Notice that, at any point $\vec x$ in the support of the source, a ball $B_{r_{eff}}(\vec x)$ should be excluded from the integration domain, because only mass elements sufficiently distant from $\vec x$ contribute to the classical potential in the known manner (\ref{eq: classical potential for spherical body derived from the modified S solution}). The underlying assumption is that - whatever the analytic continuation of the solution (\ref{eq: asymptotically Schwrzschild solutions}) may be in the strong-field regime - the potential in any small region of an extended source is always mostly determined by the mass distribution outside of it, as it happens in the purely Newtonian case. The radius $r_{eff}$ can be determined for different $n$'s by the condition $\frac{c_1}{r_{eff}^n} \ll \frac{2M_\odot G}{r_{eff}}$, in combination with the upper bounds on $c_1$ from Solar System observations \cite{Scali2024}. This automatically regularize the self-force divergence in the second integrand, which is more severe than in the Newtonian counterpart and would make the integral diverge in the ultraviolet for most of the $n$'s.\footnote{For example, assuming a regular $\rho_V$ and spherical symmetry, the integrand has a pole of order $n-2$, so that only the case $n=2$ is convergent. Also, since the integrand is positive and no angular cancellation can occur near the pole, the integral cannot even be defined in the sense of the Cauchy principal value. Assuming axial symmetry instead, the pole is of order $n-1$, so that for no $n$ the integral is convergent.}

In this context, the consistency bound derived above on the angular momentum of locally orbiting particles may translate, at most, into a bound on the gradient of the velocity field of the continuous matter distribution. This is because the study of geodesics is performed in the background (\ref{eq: asymptotically Schwrzschild solutions}), which is a local solution. In this sense, the applicability of the model for different values of $n$ would depend on the choice of extended source and the degree of its differential rotation. The constraints on $c_1$ derived in the Solar System \cite{Scali2024} could be used in this context; however, in the spirit of allowing a scale dependence of the gravitational interaction - embodied by $c_1$ - it would be consistent to provide new upper bounds on $c_1$ by fitting more appropriate observables at these scales. A possible bound on the velocity-field gradient could then be adopted as a working hypothesis, to be checked \textit{a posteriori}.

\section{Conclusions}
The Solar System is one of the most precise laboratories at our disposal for testing gravity and, as such, provides the natural foundation for developing any extension of GR. Particularly, given the success of Einstein's theory in this setting, compatibility with the predictions of GR can be used to constrain the structure of a modified gravity model.

This idea was applied in the context of $f(R)$ gravity \cite{Scali2024}. The compatibility with GR was enforced at the level of the modified Einstein equations, by looking for slight deviations from the Schwarzschild line element. Specifically, general scaling relations for the free functions in the weak-field regime made it possible to recover perturbatively a family of asymptotically Schwarzschild solutions and, in turn, to reconstruct the corresponding family of $f(R)$ Lagrangians. By extracting the key steps of the procedure, it was argued how the method can be applied to other modified gravity frameworks.

With the solutions (\ref{eq: asymptotically Schwrzschild solutions}) at hand, the motion of non-relativistic test particles was investigated. It was shown how a natural definition of the (modified) Newtonian potential from the time-time component of the metric (\ref{eq: asymptotically Schwrzschild solutions}) is only consistent for particles orbiting the star with a sufficiently small angular momentum. Finally, a general expression for the modified Newtonian potential of an extended source was derived as the convolution of individual point-particle contributions (\ref{eq: classical potential for spherical body derived from the modified S solution}) with an extended volume density, Eq.\ (\ref{eq: modified Newtonian potential generated by an extended source}). Since the solution (\ref{eq: asymptotically Schwrzschild solutions}) holds in the asymptotic region and does not account for the self-gravity of the star, the convolution integral involved in the modified part of the potential must be regularized in the ultraviolet, the minimum length being provided by the local tests \cite{Scali2024}. In this context, it is argued how the consistency bound on the angular momentum of locally orbiting particles may translate in an upper bound on the velocity field gradient of the matter distribution.

The last observations point to the intriguing perspective of applying the $f(R)$ model at galactic scales. In particular, by adopting appropriate volume density distributions \cite{Binney2011}, the modified potential generated by spiral and elliptical galaxies can be derived, and the corresponding galactic observables can be computed. The goal is to assess the observable consequences of the model for galactic dynamics, also in relation to dark matter phenomenology, and to derive additional upper bounds on $c_1$ through comparison with galactic data. This will be the subject of forthcoming work \cite{ScaliFontana2026}.

More generally, the study of galactic dynamics is part of an inductive approach to modified gravity which, starting from the foundational level of the Solar System, traces the implications of the theory through galactic scales and ultimately up to cosmological scales.

\section*{Acknowledgments}
    F.S.\ thanks L.\ Amendola, S.\ L.\ Cacciatori, M.\ Fontana and O.\ F.\ Piattella for precious comments. 

\section*{ORCID}
\noindent Federico Scali - \url{https://orcid.org/0009-0004-0637-561X}

\bibliographystyle{ws-ijmpa}
\bibliography{Bibliography.bib}
\end{document}